\documentclass{iaus}
\usepackage{graphicx}

\title[Bulges at Intermediate Redshift] 
{The Fundamental Plane of Bulges at Intermediate Redshift}

\author[MacArthur \etal]   
{Lauren A. MacArthur$^1$, Richard S. Ellis$^1$, Tommaso Treu$^2$}

\affiliation{$^1$California Institute of Technology, Pasadena CA 91125, USA
\break email: lam@astro.caltech.edu\\
[\affilskip]$^2$Department of Physics, University of California, 
Santa Barbara, CA 93106-9530, USA
}

\pubyear{2007}
\volume{245}  
\pagerange{1--4}
\date{?? and in revised form ??}
\jname{Formation and Evolution of Galaxy Bulges}
\editors{M. Bureau \& C.E. Editor, eds.}

\newcommand\apj{\textit{ApJ}}%
\newcommand\apjl{\textit{ApJL}}%
\newcommand\apjs{\textit{ApJS}}%
\newcommand\mnras{\textit{MNRAS}}%

\newcommand{\ie}{i.e.\@}

\newcommand{\magarc}{\ifmmode {{{{\rm mag}~{\rm arcsec}}^{-2}}}
             \else {{{mag}$~${arcsec}$^{-2}$}}
             \fi}

\newcommand{\apgt}{\ {\raise-.5ex\hbox{$\buildrel>\over\sim$}}\ }
\newcommand{\aplt}{\ {\raise-.5ex\hbox{$\buildrel<\over\sim$}}\ }
\newcommand{\apeq}{\ {\raise-.5ex\hbox{$\buildrel<\over=$}}\ }
\newcommand{\gtlt}{\ {\raise-.5ex\hbox{$\buildrel<\over>$}}\ }
\newcommand{\sersic}{S\'{e}rsic}
\newcommand{\hunit}
           {km\,s$^{\hbox{\scriptsize -1}}$\,Mpc$^{\hbox{\scriptsize -1}}$}
\newcommand{\hub}{H$_{\hbox{\scriptsize 0}}$}

\begin{document}

\maketitle

\begin{abstract}
We report on a new study aimed at understanding the diversity and evolutionary 
properties of distant galactic bulges in the context of well-established 
trends for pure spheroidal galaxies.  Bulges have been isolated for a sample 
of 137 spiral galaxies in the GOODS fields within the redshift range 
0.1\,$<$\,$z$\,$<$\,1.2.  Using proven photometric techniques we 
determine for each galaxy the characteristic parameters (size, surface 
brightness, profile shape) in the 4 GOODS-ACS imaging bands of both the disk 
and bulge components.  Using the DEIMOS spectrograph on Keck, precision 
stellar velocity dispersions were secured for a sizeable fraction of the 
bulges. This has enabled us to compare the Fundamental Plane of our distant 
bulges with that of field spheroidal galaxies in a similar redshift range.  
Bulges in spiral galaxies with a bulge-to-total luminosity 
ratio ($B/T$)\,$>$\,0.2 
show very similar patterns of evolution to those seen for low luminosity 
spheroidals.  To first order, their recent mass assembly histories
are equivalent. 
\end{abstract}

\firstsection 
\section{Introduction}
Significant progress has been made recently about the mass dependence of
the evolution of the field spheroidal (E/S0) galaxy population via 
Fundamental Plane (FP) analyses to $z$\,$\simeq$\,1 (\eg Treu \etal\ 2005, 
hereafter T05; van~der~Wel \etal\ 2005).  Whereas the most massive spheroidal 
galaxies support the long-held view of early collapse and subsequent passive 
evolution, a surprising amount of recent star formation (SF) is necessary to 
explain the scatter and FP offsets for lower mass galaxies.  T05 find 
that as much as 20--40\% of the present dynamical mass in systems with 
$M$\,$<\,$10$^{11}M_{\odot}$ formed since $z$\,$\simeq$\,1.2.  

Meanwhile, the evolutionary history of galactic bulges, a key issue 
in the origin of the Hubble sequence, remains poorly understood.  Originally 
thought to form at high-$z$ through dissipationless collapse, their continued 
growth, as predicted in hierarchical models, is consistent 
with the diversity observed in their present-day stellar populations.  
Local data alone, however, cannot distinguish between quite different 
hypotheses for bulge formation, including dynamical rearrangement of disk 
material triggered by interactions and the evolution of bars.  
Bulges grown in this way would show different evolutionary trends than those 
of isolated elliptical galaxies.  FP observations at a range of epochs could 
thus distinguish between the different formation scenarios.

A first glimpse at the photometric properties of bulges at intermediate 
redshift was presented by Ellis, Abraham, \& Dickinson (2001, hereafter EAD),
who examined aperture colors of bulges in 68 isolated spirals with 
$I_{AB}$\,$<$\,24 in the Hubble Deep Fields. EAD found a remarkable 
diversity in bulge colors over the redshift range 0.3\,$<$\,$z$\,$<$\,1, with 
few as red as a passively-evolving track that matches the integrated colors 
of luminous spheroidal (E/S0) galaxies.  They concluded that bulges have 
experienced recent periodic episodes of rejuvenation consistent with 15--30\% 
growth in stellar mass since $z$\,$\simeq$\,1.  These conclusions were 
challenged by Koo \etal\ (2005) who located 52 luminous ($I_{AB}$\,$<$\,24) 
bulges at $z$\,$\simeq$\,0.8 in the Groth Strip Survey for which a more 
elaborate photometric decomposition was undertaken to isolate the bulge 
component.  They found that 85\% of their field sample had uniformly red 
colors consistent with passively-evolving coeval systems. Only a 
minority (8\%) showed blue rest-frame colors of which a majority occurred in 
interacting or merging systems.  It is important to note, however, the 
differences between the samples of these two analyses: EAD select on total 
galaxy magnitude and thus sample the full Hubble sequence of spirals; 
Koo \etal\ select on {\it bulge} luminosity and thus only sample the brightest 
(massive) end of the bulge luminosity function.  If bulges indeed follow the 
same trends as a function of mass as seen in spheroidals, the early 
discrepancies are not surprising.
Here we report on progress of a study aimed at illuminating the relationship 
between bulges and spheroidals by extending the color-based studies with 
kinematics and adding galaxy bulges to the FP studies (MacArthur \etal\ 
in prep.\@).

\section{Sample \& Data}\label{sec:sample}
The galaxies for this analysis are selected from the GOODS fields for which
deep imaging in four Hubble Space Telescope Advanced Camera for Surveys 
({\it HST}-ACS) passbands are available (Giavalisco \etal\ 2004).
For our FP analysis, the spectroscopic sample is drawn from two campaigns
using the DEIMOS spectrograph on Keck; (i) that of T05, who 
secured high $S/N$ spectra of a magnitude-limited ($z_{AB}$\,$\,<\,$22.5) 
sample of isolated spheroidal galaxies in GOODS-North  
spanning 0.1\,$<$\,$z$\,$\,<\,$1.2, and (ii) a later 2005 campaign dedicated 
to increasing the spiral sample for the current project, also 
limited at $z_{AB}$\,$<$\,22.5, yielding a further target sample of 45 
spirals in GOODS within 0.1\,$<$\,$z$\,$<$\,0.7.

\begin{figure*}
\begin{center}
\includegraphics[width=0.98\textwidth]{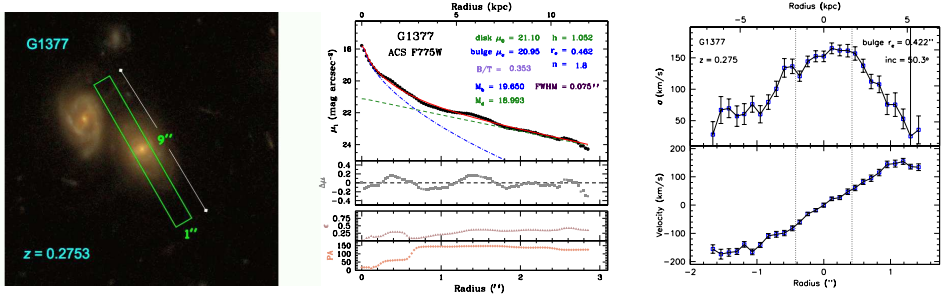}
\caption{Example of observational setup with photometric and 
         kinematic analysis for a spiral galaxy at $z$\,$\sim$\,0.3.  
         {\it Left}: GOODS {\it HST}-ACS image with DEIMOS slit position 
         indicated. {\it Middle}: ACS F775W band SB profile 
         (circles in {\it upper panels}) and bulge/disk decomposition.  
         The dashed line represents 
         the disk fit, the dash-dotted line is the bulge, and the solid line
         is the total profile fit.  The bottom three panels show the fit
         residuals, and the run of fitted isophote ellipticity and position
         angle respectively.
         {\it Right}: Measured kinematic profiles from 2D DEIMOS
         spectra. {\it Top}: velocity dispersion as a function of the
         light-weighted radius.  {\it Bottom}: radial velocity 
         profiles (\ie\ rotation curves), shifted to zero velocity at 
         $r$=0.
         The vertical dashed lines are at $\pm1\,r_e$.
         \label{fig:obs}}
\end{center}
\end{figure*}

For all galaxies, 1D surface brightness (SB) profiles are extracted from the
ACS images and modeled either as single component \sersic\ (generalized 
Gaussian) profiles or, when a disk component is evident, are decomposed 
simultaneously into bulge and disk components using the techniques outlined 
in MacArthur, Courteau, \& Holtzman (2003).  We model the disk 
light with a pure exponential function and the bulges with a \sersic\ 
profile.  An example of our B/D decomposition is shown in Fig.~\ref{fig:obs} 
[middle panel].

Velocity dispersion and rotation profiles were measured using the 
well-tested Gauss-Hermite Pixel Fitting algorithm of 
van~der~Marel (1994).  Precision central velocity dispersions, $\sigma_0$, 
were secured for 181 galaxies in T05 and for 23 spiral bulges in the second 
campaign.  An example of our derived kinematics is shown in the right panels 
of Fig.~\ref{fig:obs}.  We standardize the $\sigma$s to an aperture of $r_e/8$ 
according to the relation derived from early-type galaxy kinematic profiles 
of Cappellari \etal\ (2006).
Finally, to minimize disk 
contamination to the measured kinematics, we limit our FP sample to spheroids 
residing in galaxies with $B/T$\,$>$\,0.2, leaving us a total sample
of 56 single- and 91 two-component galaxies at 0.1\,$<$\,$z$\,$\,<\,$1.2.
Throughout we adopt a flat cosmology with $\Omega_M$\,=\,0.3, 
$\Omega_{\Lambda}$\,=\,0.7, \hub\,=\,65~\hunit, and the AB magnitude system.

\subsection{Observed Colors}
\label{sec:colors}
Regarding the diversity in observed colors for our 
sample, we find that most of the single component galaxies are red and 
consistent with a single burst model with $z_f$\,$\simeq$\,3.  On the other 
hand, the spiral bulges scatter strongly to the blue, consonant with EAD's 
findings.  Unlike EAD, however, we do see a few bulges that are at least as 
red as some of the E/S0s.  These red bulges could be akin to those
observed by Koo \etal.

The question thus arises whether we are looking at two different populations
each following a distinct formation path, or rather a more continuous mode 
of spheroid building, with the massive ellipticals generally being at a later 
stage of their evolution.  Rephrased, we can ask whether bulges 
and E/S0s {\it at a given mass} are consistent with each other in terms of 
their evolutionary paths.  To address this issue, we now consider the 
combined set of dynamical and photometric information.

\section{Fundamental Plane Evolution}
The FP is traditionally given as a relation between a galaxy's
effective radius, $R_e$, the average SB within $R_e$, SBe, and the
central velocity dispersion, $\sigma_0$.  In order to evaluate evolutionary 
trends as a function of $M/L$ ratio, it is necessary to 
convert these measured parameters into masses.  An effective dynamical mass 
can be defined using the scalar virial theorem for a stationary stellar system 
as $M \equiv c_{2}(n)\,\sigma_{0}^{2}\,R_e/G$, where $c_{2}(n)$ is the virial 
coefficient and $n$ is the \sersic\ shape parameter.  In the case of 
structural homology, the virial coefficient is a constant for 
all galaxies and the FP maps directly into a $M/L$ ratio (\eg T05).
However, in the case of varying $n$, the profile shape does affect the 
$M/L$ through a variation in the $\sigma(r)$ profile.
Several studies have constructed dynamical models to derive 
$c_2(n)$ for different profile shapes with generally consistent results.  
We use the derivation of Trujillo \etal\ (2004, hereafter Truj04), 
who developed non-rotating isotropic spherical models accounting for
the projection of $\sigma(r)$ over an effective aperture.

For the purpose of assessing evolutionary trends, we must have a local 
relation (at $z$\,=\,0) against which to compare higher-$z$ galaxies. 
As far as we are aware, the only analysis to date that provides a suitable 
local relation taking into account the varying profile shapes
is that of Truj04.  Their FP sample includes 45 local cluster ellipticals
drawn from the literature.
Converting the Truj04 relation to the units and cosmology used here 
their local relation reads,
$\log_{10}(M/L_B)_{0}$\,=\,0.091\,$\log_{10}(M)$\,+\,0.063,
and we measure the offset for galaxy $i$ as,
$\Delta\log_{10}(M/L)^i$\,=\,$\log_{10}(M/L)^i$\,$-$\,$\log_{10}(M/L)_0^i$.
\begin{figure*}
\begin{center}
\includegraphics[width=0.97\textwidth,bb=20 172 572 362]{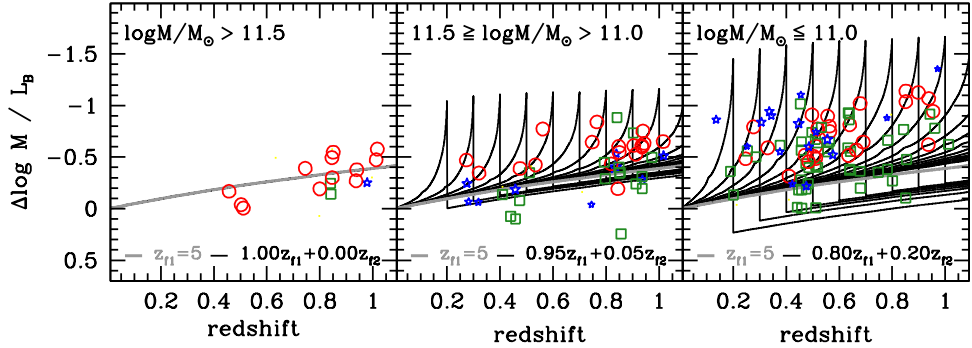}
\caption{
         Offset of $B$-band $M/L$ ratio, $\Delta\log_{10}M/L_B$ 
         versus redshift and divided into
         3 dynamical mass bins.  The gray line indicates the evolutionary
         trend for a system that formed in a single burst at $z_f$=5 and
         evolves passively based on the BC03 models.  Dark lines illustrate 
         the effects of secondary burst of star formation (5\% or 20\% by 
         mass) at $z_{f2}$\,=\,0.1,\,0.2,...
         added to the initial burst.  Circles indicate the single component
         spheroidal galaxies with Hubble T-type$<$3 (classified as in T05), 
         squares 
         are T$<$3 galaxies decomposed into disk and bulge components, and
         stars are spiral bulges with T$\ge$3 (also decomposed into 
         2-components).
         \label{fig:results}}
\end{center}
\end{figure*}
Figure~\ref{fig:results} plots $\Delta\log_{10}M/L_B$ as a function
of redshift, divided into three spheroid mass bins.  This representation
can be interpreted in terms of SF histories in an analogous
manner to T05 (Fig.~21).  Over-plotted are $M/L$ evolution 
models from Bruzual \& Charlot (2003; see caption for details).  
The most massive spheroids (including E/S0s and bulges) 
are well described by an old population formed in a single burst.  In the 
lower mass bins, however, many galaxies 
deviate significantly from this relation.  This can be explained as more 
recent spheroid building via subsequent bursts of SF on top of an
underlying old population, where the old population dominates the total stellar
mass.  The black lines in the two lower mass bins represent the $M/L$ evolution
for such models.  For the intermediate mass range,
the data are consistent with a more recent
burst of SF that represents 5\% of the total stellar mass, and in the
lowest mass bin, recent bursts involving up to 20\% of the total mass are
required.  Due to the short timescales of the initial $M/L$ decline after
a burst of SF, these mass fraction estimates are likely underestimates.
Broadly speaking, it appears that the bulges follow a similar relation
to that defined by the E/S0s at the same redshift.  In particular, there is 
no evidence of an
offset to lower $M/L$ ratios (due to increased $B$-band luminosity) which 
may be expected if the bulges are undergoing a more continuous mode of SF, 
as is expected in the secular formation scenario.  

We thus conclude that all spheroids {\it at a given mass}, and whose 
luminosity contributes at least 20\% of the total galaxy light, 
appear to follow the same evolutionary path.  Lower mass 
spheroids must have had more recent stellar mass growth, so the recent 
observations of ``downsizing'' in spheroidal galaxies (\eg Bundy \etal\ 2005) 
extends to the bulges of spiral galaxies.

\begin{acknowledgments}
We would like to thank the conference organizers for putting together
such a stimulating and informative meeting.  L.A.M acknowledges partial
financial support from NSERC.
\end{acknowledgments}


\end{document}